 \documentclass[preprint,aps]{revtex4}
 \usepackage{amssymb} \usepackage{graphicx} \graphicspath{ {images/} }
  \usepackage{dcolumn} \usepackage{amsmath}
 \usepackage{tikz}
  \usetikzlibrary{shapes,backgrounds}
\usepackage{color}

\begin{document}
 \title{New Developments in Killing Spinor Programme and More Motivations for Physics}

 \author{\"{O}zg\"{u}r A\c{c}{\i}k}
\email{ozacik@science.ankara.edu.tr}
\address{Department of Physics,
Ankara University, Faculty of Sciences, 06100, Tando\u gan-Ankara,
Turkey\\}

\begin{abstract}
This present work is based on our previous publications, which all together trigger our ''Killing spinor programme''. Other significant spinor fields are injected into the scheme and the intricate relations between them and their bilinears are constructed here. The web of relations concerning these constructions are presented clearly in a Venn diagram. Particular attention is devoted to the construction of equations satisfied by the homogeneous bilinears induced from Dirac spinors. Contacts with various physics theories are made throughout the text and some of the future directions of the programme are defined.

\end{abstract}

\maketitle

\section{Introduction}

In a recent work \cite{Acik Ertem 2015}, the geometric and physical properties of bilinear covariants corresponding to twistors (TSs) and Killings spinor (KSs) were investigated. The main objects, namely the higher degree Dirac currents \footnote{For a massive Dirac spinor, the metric dual of the degree$=1$ part exactly corresponds to the streamlines of electrons and the currents defined by the tangent vector fields of these streamlines are termed as Dirac currents in the literature.} were defined as the homogeneous parts of these spinor bilinears and for the case of Killing spinors, they were subject to some reality conditions \footnote{We also accept the name \textit{generalized Dirac currents} as synonymous to \textit{higher-degree Dirac currents} and prefer its usage throughout this work with the abreviation GDCs.}. The reality conditions were seen to arise as a consequence of Generalised Dirac Currents' (GDCs) definitive dependence on the inner product types of real or complex spinors and to other parameters of the theory which all together build up the data set for Killing Spinor Generalised Dirac Currents (KS-GDCs). The original discovery was the existence of a set of coupled first order differential equations for these KS-bilinears that gave rise to other remarkable geometric and physical equations \cite{Acik Ertem 2015 insight}. There are two kinds of sets that we call \textit{principal} and each is composed of two differential equations for the same current, one of them contains the exterior derivative and the other the coderivative of the associated homogeneous spinor bilinear. There seem to be many more consequences of these equations with a variety of applications in mathematical physics that we somehow started to deduce a few of them \cite{Acik Ertem 2016, Ertem 2016a, Ertem 2016b, Ertem 2016c}. \\

In this work we develop our \textit{Killing Spinor Programme} further so as to include other significant spinor fields and building up the intricate relations between them and their bilinears; this web of relations is presented roughly in a Venn diagram. Because of their leading role inside the programme the study of Killing spinors are delayed till to the end and other specific spinors are analysed step by step. These are twistors, Dirac spinors, Weyl spinors, parallel spinors an Killing spinors respectively. An original unifying formulation for Penrose and Dirac operators is given in the language of bundles rather clearly. Especially the generalised Dirac currents of Dirac spinors are carefully worked out and it is made explicit that the resulting equations will generate new directions for physical motivation. Before demonstrating the original part of the present work, an intention occurs for rewriting the reality conditions for spinor induced generalised Dirac currents (S-GDCs) and also reclassifying the possibilities induced from the data sets. Doing this will enable us to strengthen our language with a slight change of notation and also will shed light onto the classification scheme existent in the realm of Killing spinors and their bilinears. Analysis of reality conditions were mandatory for the Killing spinor case i.e. for KS-GDCs \footnote{In the orginal paper \cite{Acik Ertem 2015} there were some minor errors in this context.} and in contrary twistor spinor generalised Dirac currents needed no analysis for reality conditions; that was because the corresponding data sets didn't contain the inner product representatives or equivalently the adjoint involution representatives as their elements. \\

Contact with physics is scattered throughout the text which contains Dirac's 1928 theory of the electron, supersymmetry and quantum field theory, extended fields and particles, possible applications in (canonical) quantum gravity theories, curved spacetime quantum field theory and the Cauchy evolution of physical fields generated by Killing spinors, gravitation induced electrical currents in superconductors, Rarita-Schwinger field and spin gauge theories.\\

The organization of the paper is as follows. In section 2, we give some details about spinors and their bilinears in Clifford language. There we define the generalised Dirac current of an arbitrary spinor and the necessary reality conditions if required by the corresponding data set. This is done with a view in complete accordance with our previous work \cite{Acik Ertem 2015} which we call the \textit{initial approach}. The other alternative approach named the \textit{detailed approach} is not treated in this study; but some remarkable hints about it can be found in section 2. Section 3 contains the introduction of specific kinds of spinor fields with brief overviews about their properties that are used for obtaining the first order differential equations satisfied by their generalised Dirac currents. The properties about Killing spinor field induced generalised Dirac currents, which are our main objects of interest, are summarized in another subsection. Their classification scheme in terms of the Clifford expectation values of a certain algebraic operator, resulting with the principal sets of equations is given. Section 4 is devoted to studying the details of the arrows inherent in the Venn diagram followed by emphasizing the roles of co-frame fields and spinor frames in physical respects. The arrow part contains new material for the development of the Killing spinor programme that is thought of as a physical theory rather than mathematical. Section 5 concludes the paper. There are three appendices supplementary for the bulk of the paper the last of which derives the principal equations for the generalised Dirac currents of Dirac spinors.


\def\nblx#1{\nabla_{X_{#1}}}
\def\sbl#1{(\psi\overline{\psi})_{#1}}

\section{Clifford spinors and their generalised Dirac Currents}

\subsection{Clifford algebras}
The mathematical language we use for spinor calculations is based on the Clifford bundle formulation of spacetime \cite{Benn Tucker, Tucker daktilo, Charlton} and this enables us to use the advantage of taking spinor fields as smooth sections of the (local) left ideal subbundle; this definition is due to M. Riesz \cite{Riesz}. Hence a spinor at a point belongs to one of the left ideals of the fibre (Clifford) algebra generated by the cotangent space at that point, of course for the simple (irreducible) case the left ideals are minimal i.e. projected by a primitive idempotent of the algebra. Six low dimensional Clifford algebras together with the three standard periodicity relations determine the whole structure of real Clifford algebras for the finite $n$-dimensional case \footnote{In effect the structure is determined by the knowledge of the dimension and the signature of the generating space $n=n_{+}+n_{-}$ where $n_{+}(n_{-})$ is the $g$-normalized spacelike (timelike) base covectors. We adopt the multiply plus convention for the signature of spacetime as relativists do but there are some claims that the converse choice is more appropriate for physical reasons (see page 22 of \cite{Hestenes}). } and also the dimensions of the real representations carried by minimal left ideals could be calculated as $2^{\lfloor\frac{n}{2}\rfloor+\sigma}, \sigma=0,1$ \cite{Acik I}. Handling complex Clifford algebras is much easier where the knowledge of two low dimensional complex Clifford algebras and only one periodicity relation is enough for determining their algebraic structures \footnote{In the complex case signature has no meaning; so in effect, only the complex dimension $n$ of the generating space determines the whole algebraic structure of complex Clifford algebras}, the dimensions of the complex representations carried by minimal left ideals could be directly seen to be $2^{\lfloor\frac{n}{2}\rfloor}$ for any finite $n$. \\

Using the representatives of the following quantities: the projector primitive $P$, an involution of the algebra $\mathcal{J}$, an invertible element $J$ tying the primitive to the involuted primitive by a self-similarity transformation and the induced involution $j$ on the division factor algebra; then one can construct all of the $7$ main typical classes of real spinor inner products and also the remaining $3$ swap-types of reducible-spinor inner product representatives $(.,.)$ over the related left ideal classes \cite{Acik Ertem 2015}, \cite{Benn Tucker}. The set of real inner product classes contracts to a proper subset under complexification because of the dominant character of complex numbers in the set of real associative division algebras endowed with the operation $\otimes_{\mathbb{R}}$. Perhaps there are some differences between the real case and complex case apparent especially when the classification with respect to adjoint involution classes are selected. Then we define the data set of an inner product class with representative $(.,.)$ as $\Delta_{(.,.)}:=\{(n_{+},n_{-}), [P], [J],[\mathcal{J}],[j]\}$ where the last quantity is of secondary kind because of its dependence to the formers. For the complex case, signature looses its meaning and $\Delta_{(.,.)}$ becomes $\{n, [P], [J],[\mathcal{J}],[j]\}$; $(n_{+},n_{-})$ is the signature of the metric field $g$ and the dimension is $n=n_{+}+n_{-}$. \\

The algebra involution used to define the spinor inner product is called its \textit{adjoint involution}, because the induced linear transformation of the left Clifford multiplication by an element has as its adjoint linear transformation the one defined by left Clifford multiplication with the image of the same element under the involution. Two products belong to the same class if and only if the associated involutions are equivalent. As a consequence we may turn our attention to involution classes rather then the product classes in some situations and define the data set associated to the involution class represented by $\mathcal{J}$ obviously as $\Delta_{\mathcal{J}}:=\{(n_{+},n_{-}), P, J,(.,.),j\}$, this time we have used the representatives rather then the classes for brevity. This freedom of choice between inner products and adjoint involutions will be important for interpreting the results of our analysis.

\subsection{Spinors}
We treat real (semi-)spinors as minimal left ideals of real (semi-)simple Clifford algebras; let $(\frac{S}{2})S$ denote them respectively. From the Wedderburn factorisation theorem for simple associative algebras it is known that simple real Clifford algebras are isomorphic to tensor products of associative real division algebras and total matrix algebras over reals. Moreover, from the theorem of Frobenius we know that the only associative real division algebras are the real numbers $\mathbb{R}$, complex numbers $\mathbb{C}$ and quaternions $\mathbb{H}$. So we can think of a (semi-)simple Clifford algebra $(\frac{C}{2})C$ as a total matrix algebra over $\mathbb{D}$. A minimal left ideal $X$ is a left $C$-module and a right $\mathbb{D}$-module simultaneously; where $\mathbb{D}\in \{\mathbb{R},\mathbb{C},\mathbb{H}\}$ and $X\in\{S,\frac{S}{2}\}$ accordingly. Then the algebra isomorphisms known as the Fierz identities are
\begin{equation}
S_{\mathbb{D}}\otimes  {_{\mathbb{D}}}S^{*}=End_{\mathbb{D}}(S_{\mathbb{D}})\equiv {_{\mathbb{D}}}End ({_{\mathbb{D}}}S^{*})\simeq C(\mathbb{D}) \nonumber
\end{equation}
or for the reducible case
\begin{equation}
\frac{S_{\mathbb{D}}}{2}\otimes  \frac{{_{\mathbb{D}}}S^{*}}{2}=End_{\mathbb{D}}(\frac{S_{\mathbb{D}}}{2})\equiv {_{\mathbb{D}}}End (\frac{{_{\mathbb{D}}}S^{*}}{2})\simeq \frac{C(\mathbb{D})}{2}, \nonumber
\end{equation}
 note that the $\mathbb{D}$-module structures are in prominence. Here $S_{\mathbb{D}}$ is the minimal left $C$-ideal $S$ considered as a right $\mathbb{D}$-module and its dual $(S_{\mathbb{D}})^{*}={_{\mathbb{D}}}S^{*}$ is the minimal right $C$-ideal $S^{*}$ considered as a left $\mathbb{D}$-module. Furthermore $End_{\mathbb{D}}(S_{\mathbb{D}})$ is the vector space of right $\mathbb{D}$-linear transformations of $S_{\mathbb{D}}$ and ${_{\mathbb{D}}}End ({_{\mathbb{D}}}S^{*})$ the vector space of left $\mathbb{D}$-linear transformations of ${_{\mathbb{D}}}S^{*}$ which are both isomorphic to $C$ considered as a $\mathbb{D}$-linear algebra. It should also be noted that ${_{\mathbb{D}}}S^{*}$ can be thought as $L_{\mathbb{D}}(S_{\mathbb{D}},\mathbb{D})$, that is the space of $\mathbb{D}$-valued, right $\mathbb{D}$-linear maps over $S_{\mathbb{D}}$ and vice-versa, i.e. $S_{\mathbb{D}}$ as ${_{\mathbb{D}}}L({_{\mathbb{D}}}S^{*},\mathbb{D})$. Likewise all are valid for the simple constituents of the semi-simple case.
In our language the appearance of $\mathbb{D}$-structures in Fierz identities are necessary in order to fit the real dimensions of the left and right-hand sides of the isomorphisms. For example, if the real dimension of the generating space is $n$  and $\Delta n=n_{+}-n_{-}\equiv 5\,(mod 8)$ then algebra is reducible with $\mathbb{D}=\mathbb{H}$ and the $\mathbb{D}$-dimension of the semi-spinor space is $2^{\lfloor\frac{n}{2}\rfloor-1}\stackrel{n\,odd}=2^{\frac{n-1}{2}}$; so
\begin{eqnarray}
 dim_{\mathbb{R}}C = dim_{\mathbb{R}}\mathbb{D}.dim_{\mathbb{D}}C=dim_{\mathbb{R}}\mathbb{H}. 2 dim_{\mathbb{D}}\frac{C}{2}
  &=&8 dim_{\mathbb{D}}(\frac{S_{\mathbb{D}}}{2} \otimes \frac{{_{\mathbb{D}}}S^{*}}{2}) \\ \nonumber
  &=& 8 dim_{\mathbb{D}}(\frac{S_{\mathbb{D}}}{2})^{2}=8 (2^{\frac{n-1}{2}-1})^{2}=2^{n} \nonumber
\end{eqnarray}
is the real dimension of the Clifford algebra as it should. For details we refer to \cite{Acik I}.\\

Identifying a minimal left ideal of the Clifford algebra with the spinor space gives the advantage of multiplying spinors and/or dual spinors by using the Clifford product since all ideals are also subalgebras. So, if $\psi$ is a spinor and $\overline{\phi}$ a dual spinor then by Fierz
isomorphisms one can write $\psi\otimes\overline{\phi}=\psi \overline{\phi}$. The Clifford product of a spinor $\psi$ and an adjoint spinor
$\overline{\phi}$ can be written as an inhomogeneous differential form in terms of projectors $\wp_p$'s onto $p$-form components as follows
\begin{eqnarray}
\psi\bar{\phi}&=&\sum_{p=0}^n\wp_p(\psi\bar{\phi})=\sum_{p=0}^n(\phi,\,e_{I(p)}^{\xi}\psi)e^{I(p)}\\ \nonumber
&=&(\phi,\psi)+(\phi,e_i\psi)e^i+(\phi,e_{ji}\psi)e^{ij}+...+(\phi,e_{i_p...i_2\,i_1}\psi)e^{i_1
i_2...i_p}+...+ (\phi,z^{-1}\psi)z. \nonumber
\end{eqnarray}
Whereas $(.,.)$ is the associated inner product on spinors, $e^{i_1i_2...i_p}$ is defined as $e^{i_1}\wedge e^{i_2}\wedge...\wedge e^{i_p}$ determined by an arbitrary local co-frame field $\{e^{i}\}$ and $z=*1$ is the volume fixing top-form that is globally defined if $M$ is orientable. Also $1$ is the unit element, $*$ denotes the Hodge map induced by the (smooth) metric tensor field $g$ and multi-index $I(p)$ is a well-ordered $p$-tuple of indices. For a fixed value $I_{0}(p)$ the Clifford inverse of a multi-cobasis element $e^{I_{0}(p)}$ is $e^{\xi}_{I_{0}(p)}$ for any value of $p$ i.e. $e^{I_{0}(p)}e^{\xi}_{J_{0}(p)}={\delta^{I_{0}(p)}}_{J_{0}(p)}$,  thus $e^{I(p_{0})}e^{\xi}_{I(p_{0})}=\binom{n}{p_{0}}$ for a fixed value $p_0$ of $p$ and ultimately $\sum_{p=0}^{n} e^{I(p)}e^{\xi}_{I(p)}=2^{n}$ (Appendices of \cite{Acik Ertem 2015}).

Let us make the abbreviation $(\psi\bar{\phi})_{p}$ for $\wp_p(\psi\bar{\phi})$, then the $p$-form component of the \textit{coupled-bilinear} (i.e. $\psi\neq\phi$) is written as
\begin{equation}
(\psi\bar{\phi})_{p}=(\phi,e_{i_p...i_2\,i_1}\psi)e^{i_1i_2...i_p},
\end{equation}
and the \textit{singled-bilinear} (i.e. $\psi=\phi$) $p$-form component as
\begin{equation}
(\psi\bar{\psi})_{p}=(\psi,e_{i_p...i_2\,i_1}\psi)e^{i_1i_2...i_p}.
\end{equation}
Here care must be taken because in general the local spinor inner products of spinor basis are $\mathbb{D}$-valued functions over $M$! There is a special basis for spinors assigning constant values to these products; for details see the Section 4. When regarding the reality conditions of the singled-bilinears, there are two ways which we call the \textit{initial approach} and the \textit{detailed approach} respectively. Present work focuses on the initial approach and the detailed one will be treated elsewhere. The latter approach will depend on the role of complex numbers in the set of real associative division algebras for the consideration of real Clifford algebras. In the complex case the operation of tensoring with $\mathbb{C}$ and complex conjugation will play a leading role. Hence, concepts such as complex structures, complexification, realification and Cayley-Dickson process will be of importance in the so called detailed approach. \\

\textbf{Initial Approach (IA)}: We first want to comment on the method used and assumptions made in \cite{Acik Ertem 2015}. There the main investigation was based on the homogeneous parts of the spinor bilinears induced from twistors and Killing spinors; and it was found that the former ones were free from the restrictions implemented by the choice of the inner product of spinor fields. On contrary, the ones belonging to the latter type were subject to some constraints evidently. These constraints were inherited from the demand of defining real currents for physical reasons, that is because Killing spinors have a special character being both twistors and Dirac spinors \footnote{We mean spinors satisfying the massive Dirac equation. We also reserve the terminology Weyl spinors for spinors solving massless Dirac equation and Majorana spinors solving the real Dirac equation. Our terminology is differential instead of algebraic; in algebraic literature complex spinors are called Dirac spinors, chiral spinors are called Weyl and real spinors are termed Majorana spinors.} simultaneously. However the physical meanings of non-real currents for any kind of spinor should also separately be investigated as a problem of its own or as part of the planned work on detailed approach.\\

In the previous work, the data set for a KS-GDC was composed of the signature of spacetime, the Killing number $\lambda$, the representative adjoint involution $\mathcal{J}$ of the inner product class $[(.,.)]$, the induced involution $j_{c}$ on complex numbers $\mathbb{C}$ and the form degree, $p$, of the current. Consequently for a KS-GDC the data set is $\Delta_{KS-GDC}=\{(n_{+},n_{-}),\lambda, p, j_{c},\mathcal{J}\}$. The emphasis was made on $j_{c}$ of $\mathbb{C}$ rather then $j$ of $\mathbb{D}$, because $\lambda$ is a complex number that is either real or pure imaginary. By considering all the possibilities emerging from the collective behaviour of the elements of the data set we had reached the so called principal sets of equations, those that built up the core of our programme. There we naturally took $j_{c}$ as one of the real algebra involutions, that is the identity $Id_{\mathbb{C}}$ or the conjugation $^{*_{\mathbb{C}}}$ (or $^{*}$) on complex numbers (as a real algebra) and since the standard representatives of the adjoint involution classes on real spinors are $\xi$ or $\xi \eta$ and additionally $\xi*$ and $\xi \eta*$ for complex spinors; we completed our analysis by varying the data set $\{\lambda, p, j_{c},\mathcal{J}\}$ in accordance with the reality conditions. \\

Reality conditions were constructed by applying the complex conjugation involution to singled-bilinears of Killing spinors (eq. $(4)$) and observing their behaviour under this effect. Since the co-frame $1$-forms are all real whether they generate real or the complex algebra, the attention is only made to the reaction of the product $(\psi,e_{i_p...i_2\,i_1}\psi)$ to complex conjugation. So we declared that the singled-bilinear is real if
\begin{eqnarray}
(\psi,e_{i_p...i_2\,i_1}\psi)^{*}\stackrel{\mathcal{J}}=(\psi,e_{i_p...i_2\,i_1}\psi)\quad or\quad(\psi,e_{i_p...i_2\,i_1}\psi)^{*}=\varepsilon(e_{i_p...i_2\,i_1}\psi,\psi)\stackrel{\mathcal{J}}=(\psi,e_{i_p...i_2\,i_1}\psi), \nonumber
\end{eqnarray}
and imaginary if
\begin{eqnarray}
(\psi,e_{i_p...i_2\,i_1}\psi)^{*}=\varepsilon(e_{i_p...i_2\,i_1}\psi,\psi)\stackrel{\mathcal{J}}=-(\psi,e_{i_p...i_2\,i_1}\psi)\quad for \quad\varepsilon=\pm1. \nonumber
\end{eqnarray}
The symbol $\stackrel{\mathcal{J}}=$ means that the equalities are $\mathcal{J}$ dependent whereas the preceding ones $=$ are valid for all involutions. As a matter of course, the real $p$-forms defined from these singled-bilinears were named as the Killing spinor higher-degree Dirac currents independent of their real or complex character i.e. whether the spinors are imbedded in a real or a complex Clifford algebra. Note that this operation is valid for any kind of spinor whether it is a solution of a differential equation or not; consequently we can define their generalised Dirac currents with the reality conditions given above. A known thing is that $\mathbb{C}^{*}$-symmetric ($\varepsilon=+1$) inner products and $\mathbb{C}^{*}$-skew ($\varepsilon=-1$) inner products are equivalent (), it is traditional not to chose $\varepsilon=-1$ case instead of $\varepsilon=+1$, because of the existence of a useful quantifier called the (\textit{Witt}) \textit{index} for the symmetric case($\varepsilon=+1$).\\

From this \textit{initial approach} (IA), it is appearant that the reality conditions for singled-bilinears (induced from real/complex-spinors) are only apt to $\mathbb{C}^*$-symmetric inner-products; otherwise they are all termed real and no further analysis is necessary!\\

The IA-reality conditions, for any real/complex spinor generalised Dirac current ($S^{\Bbbk}-GDC$, $\Bbbk=\mathbb{R},\mathbb{C}$) admitting ${\cal{J}}=\xi\eta$ or $\xi\eta^*$ as the adjoint involution representatives, are given by the upper half below
\begin{eqnarray}
S^{\Bbbk}-GDC &=&\left\{
                \begin{array}{ll}
                ($i$\psi\bar{\psi})_{p}, & \hbox{$\lfloor p/2 \rfloor + p=odd$ ,} \\
                (\psi\bar{\psi})_{p}, & \hbox{$\lfloor p/2 \rfloor + p=even$,}\\ \hline
                ($i$\psi\bar{\psi})_{p}, & \hbox{$\lfloor p/2 \rfloor=odd$ ,} \\
                (\psi\bar{\psi})_{p}, & \hbox{$\lfloor p/2 \rfloor=even$,}
                \end{array}
                \right.
\end{eqnarray}
the ones with the adjoint involutions ${\cal{J}}\simeq\xi\eta$ or $\xi\eta^*$ are given by the lower half; upper (lower) means above (below) the horizontal line at the right hand side. For the real case there are only two separate representative involution types, $\xi$ and $\xi\eta$, and they both could be the adjoints of $\mathbb{C}^{*}$-symmetric spinor products determined by the dimension and signature. For the complex case there are also the additional types namely the ones produced by composing the formers with complex conjugation and only in these additional cases the representatives could be the adjoints of $\mathbb{C}^{*}$-symmetric inner products which are permitted by the dimension. In cases other then the $\mathbb{C}^{*}$-symmetric ones, the generalised Dirac currents are all equal to $(\psi\bar{\psi})_p$ since it is inert under complex conjugation. Clearly, the only marker for constructing reality conditions is the complex conjugation when IA is in use. Notice that in real Clifford algebras we can obtain purely imaginary $p$-form fields from real spinor fields because of the hereditary features of spinor inner products.

\section{Specific spinor fields in curved spacetimes}

\subsection{Specific spinors}

Our knowledge about spinors and their bilinears will now be specialized by constraining spinors to the solutions of some first order differential equations. After defining the specific spinor fields by the differential equations that they satisfy, in some cases the integrability conditions deduced from the definitive differential equations will be given for their vital role in determining the necessary geometrical and topological properties that ought to be carried by the ambient spacetime. Then we will give the induced differential equations for their bilinears. The specific types of spinor fields in our interest will be twistors, Dirac spinors, Weyl spinors, Killing spinors and parallel spinors; a relational Venn diagram for these is shown in Fig. 1.
 \def\setA{(0.8,0.5) ellipse (1.2cm and 0.9cm)}
 \colorlet{ellipse edge}{blue!50}
 \colorlet{ellipse area}{blue!20}
  \def\setB{(-1.4,0.5) ellipse (2.1cm and 1.4cm)}
  \colorlet{ellipse edge}{red!50}
  \colorlet{ellipse area}{red!20}
  \def\setC{(0.8,0.5) ellipse (1.6cm and 2.4cm)}
  \colorlet{ellipse edge}{green!50}
  \colorlet{ellipse area}{green!20}
  \def\setD{(-4.1,-3.4cm) rectangle (3.9,4.7)}
  \tikzset{filled/.style={fill=ellipse area, draw=ellipse edge, thick}, outline/.style={draw=ellipse edge, thick}}
    \begin{center}
    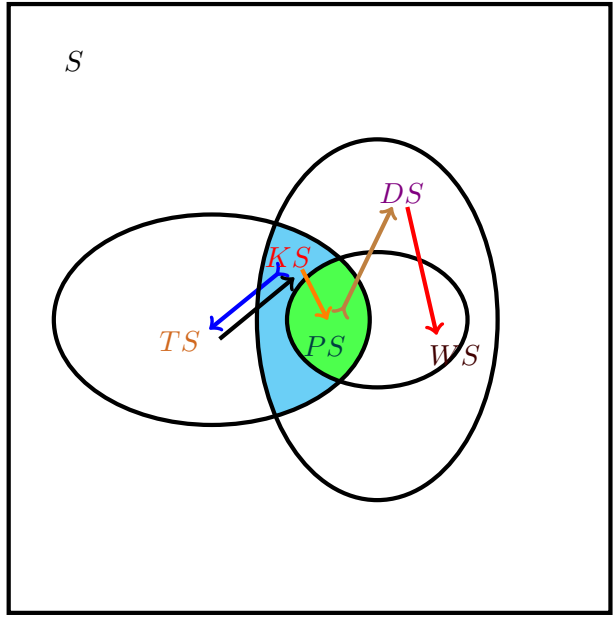
\begin{figure}
     \begin{tikzpicture}[ultra thick]
     \draw \setD;
      \begin{scope}
    \clip \setB;
    \fill[cyan!50] \setC;
      \end{scope}
      \begin{scope}
    \clip \setA;
    \fill[green!70] \setB;
      \end{scope}
       \draw \setA ;
      \draw \setB node [text=black,below left] {$\textcolor[rgb]{0.82,0.41,0.12}{TS}$};
      {\draw [<-<][blue] (-1.432,0.372) -- (-0.44,1.18); }
      {\draw [->][black] (-1.296,0.244) -- (-0.304,1.06); }
      \draw \setC ;
      {\draw [>->][brown] (0.30,0.55) -- (1,2); }
      {\node[xshift=2.8,yshift=4.4] {$\textcolor[rgb]{0.00,0.25,0.25}{PS}$} ; }
      \node[xshift=32,yshift=62] {$\textcolor[rgb]{0.50,0.00,0.50}{DS}$};
      {\draw [->][orange] (-0.20,1.17) -- (0.14,0.50); }
      {\node[xshift=-11,yshift=38] {$\textcolor[rgb]{1.00,0.00,0.00}{KS}$} ; }
      {\node[xshift=52,yshift=1] {$\textcolor[rgb]{0.25,0.00,0.00}{WS}$} ; }
      {\draw [->][red] (1.2,2) -- (1.59,0.3); }
       {\node[xshift=-92,yshift=112] {$S$} ; }
    \end{tikzpicture}
   \caption{Diagram for specific spinors}.
 \end{figure}
 \end{center}
The union of blue and green domains represent the space of Killing spinors and the green domain lonely represents the subspace containing Parallel spinors. Other spinor spaces are bounded by the smooth ellipses encircling them; i.e. twistor spinors by the big horizontal ellipse, Dirac spinors by the big vertical ellipse and Weyl spinors by the horizontal smal ellipse contained bt the former. The order for investigation indicated by this figure is from general to special, whereas we reserve the last position for the analysis of Killing spinor case. That is because our investigations suggest that they are the most remarkable ones both for physical and mathematical reasons and this claim is the central principle of our programme. Some results supporting this claim will be derived or explained during this work. The other remaining and unexplored results are hoped to be the subject of other works; for example the restrictive enforcement of the existence of Killing spinors over the geometry of spacetime is a problem of its own that we have to dispense in some way. This apprehension is inevitable if one is intending to give Killing spinors a leading role physically; a step in this direction can be taken by using distributional Ricci (scalar) curvatures coherent with the physical meaning of Killing numbers that they contain.  \\

Unless otherwise stated we do not specify whether the spinors are real or complex ($\Rightarrow$ we use $S$ rather then $S^{\Bbbk}$). For example we define the Dirac spinors those of which satisfy the Dirac equation as occuring in physics, but real spinors satisfying the same equation are called Majorana (or Majorana-Dirac) spinors in the literature. The arrows shown in the figure have meanings, while the ones with flat tails signal the existence of a contraction by the direction of the arrow together with the dimensions of the spaces at both sides, the arrows with swallow tails, namely the extensions/constructions give information about the analytical methods for building up spinors contained in the domain of the arrow tip from the spinors belonging to the domain of the arrow tail. Contractions are only possible in the presence of free parameters. Although we intend not delve into the strict details of Fig. I, we will make clear what is meant by this picture in section IV.

\subsection{Twistors and Dirac spinors}
A twistor is a solution of the equation
\begin{eqnarray}
\nabla_{X}\psi=\frac{1}{n}\widetilde{X}.\displaystyle{\not}D\psi
\end{eqnarray}
called the twistor equation, that is also described as a spinor belonging to the kernel of the Penrose operator $\mathcal{P}$ and a Dirac spinor satisfies the Dirac equation
\begin{eqnarray}
\displaystyle{\not}D\psi=m \psi.
\end{eqnarray}
An important integrability condition for twistors is given by the weaker differential equation $$\nabla_{X}({\displaystyle{\not}D}\psi)-\frac{n}{2}K_a(X)e^a.\psi=0,$$ here $K_a=\frac{1}{(n-2)}\bigg(\frac{\mathcal{R}}{2(n-1)}e_a-P_a\bigg)$ is the Schouten-Rho $1$-form \cite{Ertem 2016c}.
If $g$ is the metric tensor, $\widetilde{X}$ the $g$-dual of the vector field $X$, $n$ the dimension of the spin manifold ($n\geq 3$) and $\nabla$ the spinor connection then the Dirac operator and the Penrose operator are defined respectively as
\begin{eqnarray}
\displaystyle{\not}D\psi=e^{a}.\nabla_{X_{a}}\psi,
\end{eqnarray}
\begin{eqnarray}
\mathcal{P}\psi=e^{a}\otimes \mathcal{P}_{X_{a}}\psi \qquad;\qquad \mathcal{P}_{X_{a}}=\nabla_{X_{a}}\psi-\frac{1}{n}e_{a}.\displaystyle{\not}D\psi.
\end{eqnarray}

\textbf{A Rigorous Derivation:} From a more technical point, it will be made explicit that there are two natural first order operators over spinor fields, which of course are the above ones. Now things will become apparent immediately by the following original analysis. The starting point is based on the concept of spinor valued $1$-forms or equivalently (because of the isomorphism induced from $g$-duality) spinor valued vector fields where the first seems more natural for our notational taste \footnote{In principle we distinguish between the spinor valued $1$-forms from those used for describing higher spin fields in physics.}. These objects respectively belong to the local smooth sections of the bundles $T^{*}M\otimes S$ and $TM\otimes S$. If $\mathcal{E}$ is a bundle then we denote its smooth sections by $\Gamma\mathcal{E}$, so the $g$-isomorphism is
\begin{eqnarray}
\Gamma(T^{*}M\otimes S)\stackrel{\text{g}}{\simeq}\Gamma(TM\otimes S). \nonumber
\end{eqnarray}
The Clifford contraction map $\boldsymbol\mu$ is defined as
\begin{eqnarray}
{\boldsymbol\mu}:T^{*}M \otimes S \longrightarrow S \nonumber
\end{eqnarray}
and the induced map on sections is
\begin{eqnarray}
\widehat{\boldsymbol\mu}:\Gamma(T^{*}M \otimes S) &\longrightarrow& \Gamma S \\
\widetilde{X} \otimes \psi &\mapsto&  \widetilde{X}.\psi\:. \nonumber
\end{eqnarray}
This mapping transforms a spinor valued $1$-form in to a spinor and hence reduces the covariant degree by $1$. This is analogous to the operation of tensor contraction which reduces both the covariant and contravariant degrees by $1$ simultaneously. For this reason we call it \textit{Clifford contraction} instead of Clifford multiplication, and also it can happen that Clifford contraction can always be zero for a spinor valued $1$-form composed of a $g$-dualized vector field and a non-zero spinor but in the case of left Clifford multiplication of a spinor by a $g$-dualized vector field, vanishing of the product for every vector fields necessitates the spinor equals zero! \\

The bundle $T^{*}M \otimes S$ admits a decomposition $\Upsilon_{1}\oplus \Upsilon_{2}$ with sub-bundle projections
\begin{eqnarray}
pr_{i}: T^{*}M \otimes S=\Upsilon_{1}\oplus \Upsilon_{2} \longrightarrow\Upsilon_{i}\qquad i=1,2 \nonumber
\end{eqnarray}
such that $pr_{1}+ pr_{2}=Id_{T^{*}M \otimes S}$ where
\begin{eqnarray}
\widehat{pr}_{1}(\widetilde{X} \otimes \psi)= \frac{1}{n}e^{a}\otimes (e_{a}.\widetilde{X}.\psi) \Longrightarrow \widehat{\boldsymbol\mu}(\widehat{pr}_{1}(\widetilde{X} \otimes \psi))=\widetilde{X}.\psi \nonumber
\end{eqnarray}
and
\begin{eqnarray}
\widehat{pr}_{2}(\widetilde{X} \otimes \psi)= \widetilde{X} \otimes \psi-\frac{1}{n}e^{a}\otimes (e_{a}.\widetilde{X}.\psi)=e^{a}\otimes \widehat{\psi}_{a} \Longrightarrow \widehat{\boldsymbol\mu}(\widehat{pr}_{2}(\widetilde{X} \otimes \psi))=0. \nonumber
\end{eqnarray}
With a little algebra, it can be found that $\widehat{\psi}_{a}=\frac{1}{2n}((n-2)e_{a}.\widetilde{X}+n \widetilde{X}.e_{a}).\psi$ (see Appendix B).\\

It is clear from $\boldsymbol\mu\circ pr_{i}=\delta_{i1}\boldsymbol\mu$ that $\Upsilon_{2}=ker({\boldsymbol\mu})$ and since the spinor bundle is a left Clifford module ${\boldsymbol\mu}(\Upsilon_{1})=S$. The last equality can equivalently be written as $\Upsilon_{1}\stackrel{\boldsymbol\mu}{\simeq}S$ i.e. ${\boldsymbol\mu}|_{\Upsilon_{1}}$ is a bundle isomorphism. $\Upsilon_{2}$ subbundle is termed as the \textit{twistor bundle} and the correspondence between its smooth sections and spinor fields will follow from the decomposition of the spin connection $\nabla$ with respect to the induced projectors $\{\widehat{pr}_{i}\}$ on the sections. Let us define the operators
\begin{eqnarray}
D_{i}:\Gamma S \stackrel{\nabla}{\longrightarrow} \Gamma(T^{*}M \otimes S) \stackrel{\widehat{pr}_{i}}{\longrightarrow} \Gamma \Upsilon_{i}  \nonumber
\end{eqnarray}
so $\nabla=Id_{\Gamma(T^{*}M \otimes S)}(\nabla)=\widehat{pr}_{1}(\nabla)+\widehat{pr}_{2}(\nabla)=D_{1}+D_{2}$. Remembering $\nabla$ equals $e^{a}\otimes\nabla_{X_{a}}$ then $D_{i}\psi=\widehat{pr}_{i}(\nabla)\psi=\widehat{pr}_{i}(e^{a}\otimes\nabla_{X_{a}})\psi$. For $i=1$
\begin{eqnarray}
D_{1}\psi=\frac{1}{n}e^{b}\otimes e_{b}.\displaystyle{\not}D\psi \in \Gamma \Upsilon_{1}  \nonumber
\end{eqnarray}
and $i=2$
\begin{eqnarray}
D_{2}\psi=e^{b}\otimes\nabla_{X_{b}}\psi-\frac{1}{n}e^{b}\otimes e_{b}.\displaystyle{\not}D\psi \in \Gamma \Upsilon_{2}, \nonumber
\end{eqnarray}
i.e. $D_{2}=\mathcal{P}$. Finally, in order to pass to spinor fields; for $i=1$ we can immediately use the isomorphism property of $\widehat{\boldsymbol\mu}$ valid only in this case and $\widehat{\boldsymbol\mu}D_{1}\psi=\displaystyle{\not}D\psi$, but since $\widehat{\boldsymbol\mu}D_{2}=0$ we need an other isomorphism for passing from $\Gamma \Upsilon_{2}$ to $\Gamma S$. The most natural candidate for this comes from the right factor $\pi_{X_a}:=\pi_{a}$ of $Id_{\Gamma(T^{*}M \otimes S)}=e^ {a}\otimes \pi_{a}$, for if $\widetilde{X}\otimes \varphi \in \Gamma(T^{*}M \otimes S)$ then $\pi_{a}(\widetilde{X}\otimes \varphi)=\widetilde{X}(X_{a})\varphi$. By using this we get $\pi_{a}(D_{2}\psi)=\pi_{a}(\mathcal{P}\psi)=\mathcal{P}_{X_{a}}\psi$ for all $a$. So the origin of the significance of the Dirac and Penrose operators are explicitly shown and one has
\begin{eqnarray}
\widehat{\boldsymbol\mu}\circ\widehat{pr}_{1}\circ \nabla=\displaystyle{\not}D\quad and \quad \pi_{a}\circ\widehat{pr}_{2}\circ \nabla=\mathcal{P}_{X_{a}} \; \forall a.
\end{eqnarray}
We can also write $\widehat{\boldsymbol\mu}\circ \nabla=\displaystyle{\not}D$ because $\widehat{\boldsymbol\mu}\circ \widehat{pr}_{i}=\delta_{i1}\widehat{\boldsymbol\mu}$ or explicitly $\widehat{\boldsymbol\mu}(\nabla)=\widehat{\boldsymbol\mu}(e^{a}\otimes\nabla_{X_{a}})=e^{a}.\nabla_{X_{a}}=\displaystyle{\not}D$ and lastly $\widehat{pr}_{2}(\nabla)=\mathcal{P}$. \\

Next we give the equations satisfied by the generalised Dirac currents of twistors obtained in \cite{Acik Ertem 2015} as
\begin{eqnarray}
n d(\psi\overline{\psi})_{p}&=&(p+1)({\displaystyle{\not}d}(\psi\overline{\psi}))_{p+1}-2(p+1)i_{X^a}(\psi \overline{\nabla_{X_a}\psi})_{p+2} \\
n \delta(\psi\overline{\psi})_{p}&=&(-n+p-1)({\displaystyle{\not}d}(\psi\overline{\psi}))_{p-1}-2(-n+p-1)e^{a}\wedge(\psi \overline{\nabla_{X_a}\psi})_{p-2}
\end{eqnarray}
and for Dirac spinors (see Appendix C)
\begin{eqnarray}
(2-\Delta_{-})d(\psi \overline{\psi})_{p}&=&(\mathbf{m}-m\Delta_{-})(\psi \overline{\psi})_{p+1}-\Delta_{-} \delta(\psi \overline{\psi})_{p+2}  \\
-(2-\Delta_{+})\delta(\psi \overline{\psi})_{p}&=&(\mathbf{m}+m\Delta_{+})(\psi \overline{\psi})_{p-1}+\Delta_{+}d(\psi \overline{\psi})_{p-2}
\end{eqnarray}
where $\mathbf{m}(p,j_{\Bbbk}):=m+(-1)^{p-1} \delta(\mathcal{J}) m^{j_{\Bbbk}}$, $\Delta_{\pm}(p,\mathcal{J}):=(1\pm(-1)^{p+1}\varepsilon\delta(\mathcal{J})\mathcal{J})$ and $(e^{a})^{\mathcal{J}} =\delta(\mathcal{J})e^{a}$.

\subsection{Weyl and Parallel spinors}
Weyl spinors (or Harmonic spinors) are the complex or real (Majorona-Weyl) solutions 0of
\begin{eqnarray}
\displaystyle{\not}D\psi=0
\end{eqnarray}
and describe the massless spin-$1/2$ fields in general spacetimes. Weyl objects (spinors or bilinears) can be thought as the deformations of Dirac objects, in this case the deformation parameter is the mass of the field. Since we postpone the investigation of Killing spinors and their bilinears to the end of this section, one of the alternative ways for constructing parallel spinors is intersecting the spaces of twistors and Weyl spinors: $TS \cap WS=PS$. Then from equations $(6)$ and $(16)$, parallel spinors are the solutions of
\begin{eqnarray}
\nabla_{X}\psi=0.
\end{eqnarray}
Parallel spinors could also be defined as the deformations of Killing spinors (the view which we shall take later on). However up to now the main two sectors of specific spinor fields were logically taken as twistors and Dirac spinors; because we can build other types of spinors from them: this we call \textit{top to bottom construction}. The \textit{bottom to top construction} is based on the primary role played by Killing spinors and suits to our programme more tightly!\\

From (14)-(15) we can see that the generalised Dirac currents of Weyl spinors do satisfy
\begin{eqnarray}
(2-\Delta_{-})d(\psi \overline{\psi})_{p}&=&-\Delta_{-} \delta(\psi\overline{\psi})_{p+2},  \\
-(2-\Delta_{+})\delta(\psi \overline{\psi})_{p}&=&\Delta_{+}d(\psi\overline{\psi})_{p-2},
\end{eqnarray}
which of course imply that either $d(\psi \overline{\psi})_{p}=0$ or $\delta(\psi \overline{\psi})_{p}=0$ because if $\Delta_{\widehat{\epsilon}}=0$ then $\Delta_{-\widehat{\epsilon}}=2$; here $\epsilon$ is either $+1$ or $-1$ and should not be confused with the \textit{symmetry indicator} $\varepsilon$ of the inner-product which is also $+1$ or $-1$, and $\widehat{\epsilon}$ is defined as being the sign of $\epsilon$ i.e. $+$ or $-$.   These properties of WS-GDCs together with (12) (or equivalently with (13)) ensure
\begin{eqnarray}
(p+1)\delta(\psi \overline{\psi})_{p+2}=(-n+p+1)d(\psi \overline{\psi})_{p}=0\,,
\end{eqnarray}
satisfied by the parallel spinor generalised Dirac currents. PS-GDCs are of course closed and co-closed since they are parallel; but these equations will have implications when considered from the Killing spinor based motivation (Appendix C of \cite{Acik Ertem 2016}).

\subsection{Killing spinors}

Our main geometric and physical entities are the ones constructed out of Killing spinors; hence a review of our results for KS-GDCs will be given in this subsection. Killing spinors are defined as the solution of
\begin{eqnarray}
\nabla_{X}\psi=\lambda \widetilde{X}.\psi
\end{eqnarray}
where $\lambda \in \mathbb{C}$ termed the Killing number which is either real or pure imaginary because of the geometric constraint
\begin{eqnarray}
\mathcal{R}=-4\lambda^{2}n(n-1) \nonumber
\end{eqnarray}
imposed by the existence of such spinor fields in spacetime $(M,g)$ \cite{Baum, Leitner}. If $\lambda$ is real (imaginary) then the solution is called a real (imaginary) Killing spinor independent of the number field $\Bbbk$ of $\Gamma S^{\Bbbk}$ \cite{Leitner} (see also the last paragraph of this subsection).
Since General Relativity requires Lorentzian $4$-geometries, we assume vanishing torsion and consider pseudo-Riemannian $n$-geometries in order not to exclude Supergravity inspired torsionless geometries  \cite{Alekseevsky Cortes, Baum Friedrich Grunewald Kath, Baum, Leitner, Townsend, Penrose Rindler}. An equation that includes one of the integrability conditions for Killing spinors as a special case is $$\nabla_{X}({\displaystyle{\not}D}\psi)+\frac{\mathcal{R}}{4(n-1)}\widetilde{X}.\psi=0$$ is also known as the (E)-equation of Lichnerowicz \cite{Baum Friedrich Grunewald Kath}.\\

 By a specific choice of local spinor basis it is easy to prove the compatibility of spinorial covariant derivative with any inner product on spinor fields and hence we can pass to the corresponding differential equation on a smooth local section of a minimal right ideal bundle as $$\overline{\nabla_{X_a}\psi}=\lambda^{j_{c}}\, \overline{\psi}\, (e^a)^{\mathcal{J}}$$ and reach the important equation
\begin{eqnarray}
\nabla_{\widetilde{e_a}} \sbl p= (\hat{\lambda}^{-}_{(p)} e_a) \wedge \sbl {p-1} + i_{\widetilde{\hat{\lambda}^{+}_{(p)}e_a}} \sbl {p+1}.
\end{eqnarray}
Here the \textit{Clifford algebraic operators} are defined as $$\hat{\lambda}^{\pm}_{(p)}=(\lambda 1 \pm (-1)^{p} \lambda^{j_c} \mathcal{J})$$
with the main properties $$\hat{\lambda}^{\pm}_{(p)}e_a=\langle\hat{\lambda}^{\pm}_{(p)}\rangle e_a\qquad ; \qquad \langle\hat{\lambda}^{\pm}_{(p)}\rangle=\frac{e^{b}\hat{\lambda}^{\pm}_{(p)}e_b}{e^{b}e_b}\in \{0,2\lambda\}.$$
We recall that the local co-frame field $\{e^a\}$ should not necessarily be $g$-orthonormalised and if $\{X_a\}$ is the local frame field $\widetilde{g_{ab}e^b}=\widetilde{e_a}=X_a$. Since the Killing number is real or pure imaginary then $(\lambda^{j_c})^2=\lambda^{2}$ and hence $(\hat{\lambda}^{\pm}_{(p)})^2=2\lambda \,\hat{\lambda}^{\pm}_{(p)}$.\\

 \textbf{Quantum Considerations:} We want to comment on the quantum mechanical essence contained in the above equations; with the exception of signs coming in reverse order, the fusion of the Clifford algebraic operators with $e^a \wedge$ and $i_{\widetilde{e_a}}$ gives much physical intuition in the context of curved spacetime quantum field theory and supersymmetry. This is immediate if one remembers the interpretation of the latter operations as \textit{fermion creation} and \textit{fermion annihilation} operators respectively in \cite{E. Witten}; which together with the (bosonic) analogy given in \cite{Acik Ertem 2015} causes other reinterpretations to shoot forth. Roughly these could be built on the following observation. In (22) there at first sight exist three homogeneous neighbouring degree exterior fields and these can be interpreted as properties associated to extended fields (p-form fields, smooth fluid of p-branes) \cite{Mukherjee Tucker} or extended particles (individual p-branes, dust of p-branes) or a mixture of both, according to the details of the model \cite{Onder Tucker, Onder Tucker 2, Tucker mem, Hartley Onder Tucker}. So equation (22) may be interpreted as follows: The propagation of a brane in spacetime is accompanied by the annihilation of a brane with one higher dimension and/or by the creation of a brane with one lower dimension; of course in contrast to \cite{E. Witten} our branes belong to the bosonic sector \cite{Acik Ertem 2015 insight}. \\

 It is also of importance to note that in the case of pure field approach, one is restricted to the machinery imposed by the smooth differential geometry of classical immersions (a kind of generalised Faraday philosophy) and the particle approach signals the necessity of employing distributional differential geometry of singular immersions \cite{Dieudonne I, Hormander, Israel, Kuchar, Ibragimov, Tucker, Poisson} or the improved techniques of geometric measure theory \cite{Morgan,Federer}. One of the other modern alternative techniques for the latter case may be found in \cite{B. Yen Chen}.\\

The loop approach to quantum gravity might also be a domain of application in at least two respects 1) quantum mechanical motions of null shells therein \cite{Gambini Pullin} and 2) the relation to the quantum structure of space (spacetime) based on the concept of spin networks (spin foams) \cite{Baez, Rovelli Vidotto}. The second one could be a nice three (four) dimensional application since the cellular complex associated to a Killing spinor seems likely to fit into these network-foam schemes. In reality one can infer from the below tables that eq. (22) splits into two simultaneous equations
\begin{equation}
\nabla_{X_a} \sbl p= 2\lambda\, e_a \wedge \sbl {p-1}
\end{equation}
\begin{equation}
\nabla_{X_a} \sbl {p_{*}}= 2\lambda\, i_{X_a} \sbl {p_{*}+1}.
\end{equation}
where $p$ and $p_*$ have different parities. The lower degree special cases of these equations have given rise to conditions for closing Killing superalgebras (Appendix C of \cite{Acik Ertem 2016}). For a fixed choice of $\lambda$ these equations correspond to a row of one of the tables. For consistency, when the representative of the involution classes and the induced involution on $\mathbb{C}$ changes, $p$ should change its parity according to the red pattern whereas $p_*$ is related to the blue pattern. The lower degree equations corresponding to point particles are related to geodesics and conserved quantities along them, so the higher degree ones should be related to minimal submanifolds traced out by higher dimensional particles moving in spacetime. The homogeneous bilinears are from one side associated to gravitational charges localised on $p$-branes \cite{Acik Ertem Onder Vercin2}; if it can be shown on the other side that they calibrate the world-immersion of the brane as well, then it will be proved that these charges contribute to the inertial motion of extended particles or they may be the unique source for their inertia: a kind of classical mechanism for mass generation.
\begin{table}[h]
\centering
\begin{tabular}{|c|c|c|}
  \hline
  $     $      & $p\,odd$ & $p\,even$ \\ \hline
  $\{j_c,\mathcal{J}\}$      & $\langle\hat{\lambda}^+_{(p)}\rangle\quad\langle\hat{\lambda}^-_{(p)}\rangle$ & $\langle\hat{\lambda}^+_{(p)}\rangle\quad\langle\hat{\lambda}^-_{(p)}\rangle$ \\ \hline

  $\{Id,\xi\},\{Id,\xi^*\}$      & $\textcolor[rgb]{1.00,0.00,0.00}{0\qquad 2\lambda}$ & $\textcolor[rgb]{0.00,0.07,1.00}{2\lambda\qquad 0}$ \\
 $\{Id,\xi\eta\},\{Id,\xi\eta^*\}$      & $\textcolor[rgb]{0.00,0.07,1.00}{2\lambda\qquad 0}$ & $\textcolor[rgb]{1.00,0.00,0.00}{0\qquad 2\lambda}$ \\
  $\{*,\xi\},\{*,\xi^*\}$      & $\textcolor[rgb]{1.00,0.00,0.00}{0\qquad 2\lambda}$ & $\textcolor[rgb]{0.00,0.07,1.00}{2\lambda\qquad 0}$ \\
  $\{*,\xi\eta\},\{*,\xi\eta^*\}$      & $\textcolor[rgb]{0.00,0.07,1.00}{2\lambda\qquad 0}$ & $\textcolor[rgb]{1.00,0.00,0.00}{0\qquad 2\lambda}$ \\
  \hline
\end{tabular}
\caption{Clifford expectation values for $\lambda^{*}=\lambda$.}
\centering
\begin{tabular}{|c|c|c|}
  \hline
  $     $      & $p\,odd$ & $p\,even$ \\ \hline
  $\{j_c,\mathcal{J}\}$      & $\langle\hat{\lambda}^+_{(p)}\rangle\quad\langle\hat{\lambda}^-_{(p)}\rangle$ & $\langle\hat{\lambda}^+_{(p)}\rangle\quad\langle\hat{\lambda}^-_{(p)}\rangle$ \\ \hline

  $\{Id,\xi\},\{Id,\xi^*\}$      & $\underline{\textcolor[rgb]{1.00,0.00,0.00}{0\qquad 2\lambda}}$ & $\underline{\textcolor[rgb]{0.00,0.07,1.00}{2\lambda\qquad 0}}$ \\
 $\{Id,\xi\eta\},\{Id,\xi\eta^*\}$      & $\underline{\textcolor[rgb]{0.00,0.07,1.00}{2\lambda\qquad 0}}$ & $\underline{\textcolor[rgb]{1.00,0.00,0.00}{0\qquad 2\lambda}}$ \\
  $\{*,\xi\},\{*,\xi^*\}$      & $\textcolor[rgb]{0.00,0.07,1.00}{2\lambda\qquad 0}$ & $\textcolor[rgb]{1.00,0.00,0.00}{0\qquad 2\lambda}$ \\
  $\{*,\xi\eta\},\{*,\xi\eta^*\}$      & $\textcolor[rgb]{1.00,0.00,0.00}{0\qquad 2\lambda}$ & $\textcolor[rgb]{0.00,0.07,1.00}{2\lambda\qquad 0}$ \\
  \hline
\end{tabular}
\caption{Clifford expectation values for $\lambda^{*}=-\lambda$.}
\end{table}

The black lines under the expectation values in the second table mean their exclusion because of the incompatibility of their data sets with reality conditions. Finally from the pseudo-Riemannian identities $d=e^a \wedge \nabla_{X_a}$ and $\delta=-i_{X^a} \nabla_{X_a}$ , the differential equations satisfied by the generalised Dirac currents of Killing spinors are obtained as
\begin{equation}
d \sbl {p}=0 \qquad,\qquad \delta\sbl {p}=-2\lambda (n-p+1) \sbl {p-1}\:;
\end{equation}
\begin{equation}
d\sbl {p_{*}}= 2\lambda (p_{*}+1) \sbl {p_{*}+1}\qquad,\qquad \delta\sbl {p_{*}}=0.
\end{equation}
These are the aforementioned basic equations of our programme and some of their geometrical and physical properties were given in \cite{Acik Ertem 2015}. It was found there that these equations can be used as building blocks of Killing-Yano forms/closed conformal Killing-Yano forms, Maxwell-like equations and Petiau-Kemmer-Duffin equations. Also supergravity applications were considered in their most primitive form. Now there are other works in progress on the constructive character of KS-GDCs for determining other physical fields and their simplifying role in defining the Cauchy evolution of the data for the induced physical fields in curved spacetimes. Some further non-technical details about the Cauchy problem will be discussed in Part A of the next section. \\

Here an important point about the above tables is as follows. Although the $\{j_c,\mathcal{J}\}$ and $\{j_c,\mathcal{J}*\}$ pairs give the same results in our calculations one should be careful about the distinction between them. We took $\Bbbk$ unrestrictively in our (initial approach) considerations but since our main concern was the $\mathbb{C}^*$-symmetric (Hermitian) involutions, this requires that the above pairs must differ considerably. For example when $\Bbbk=\mathbb{R}$ then $\mathcal{J}\simeq\mathcal{J} *$ and the dimension and the signature determine whether $\mathcal{J}$ is the adjoint involution of a $\mathbb{C}^*$-symmetric spinor-inner product or not; whereas if $\Bbbk=\mathbb{C}$, $\mathcal{J}\ncong\mathcal{J} *$ and it can only be possible that $\mathcal{J}$ is the adjoint of a $\mathbb{C}^*$-symmetric spinor-inner product only if $\mathcal{J}\simeq \xi*$ or $\mathcal{J}\simeq \xi\eta*$ (see pages 78-9 and 85 of \cite{Benn Tucker}).\\

The search for spinor induced eigen-forms of Laplace-Beltrami operator (Hodge Laplacian), especially Harmonic current-forms, seems invaluable in many respects. The unvalidity of Hodge's Theorem \cite{Dieudonne II} for the indefinite case is a thing that has to be put in mind through these considerations. The coordinate-wise role of these eigenvalues particulary (in three dimensions) for the superspace of classical geometrodynamics underlines the value of the eigen-forms of the Laplace-Beltrami operator \cite{Wheeler}. More modern approaches use the eigenvalues of the Dirac operator instead of the Laplacian for the determination of the topology and geometry of the manifold under consideration \cite{Lawson Michelson}.

\section{on arrows and frames}
In this section we have two aims: 1) Commenting on the meanings of the colored arrows shown in the \textit{Diagram for specific spinors}, where some of these comments will be based on new (unpublished) results with future motivations. 2) Emphasizing the role of a special frame for spinors and paying more physical attention to it then usual. The latter will be discussed in comparison with several useful coframes defined in the differential geometric context of general relativity.

\subsection{Arrows}
The diagram in Fig. I can be thought as a primitive chart for relating the spinor types with each other and for constructing one from the other when possible. As mentioned in the paragraph after Fig.I, the flat tailed arrows refer to contractions and the swallow tailed ones to extensions/constructions.

The red, orange and black arrows are the deformations of Dirac spinors into Weyl spinors, Killing spinors into Parallel spinors and Twistors into Killing spinors respectively. The first two contractions use the inertial mass $m$ and the Killing number $\lambda$ as deformation parameters where the last is a more trickier one relying on the conformal deformations of twistors into Killing spinors (for the Riemannian case see \cite{Friedrich 1990, Schoen, Kuhnel Rademacher}). Obtaining Killing spinors from parallel spinors is also possible via a method called the \textit{cone construction} \cite{Bar} which we did not put a related arrow in the diagram because of its conceptual difference with our approach. In the literature Killing numbers are generally normalized to a rational number, but we prefer to preserve its freedom just as to qualify it as a meaningful physical parameter \cite{Acik III}.\\

The blue arrow signals a new method for building up a twistor $\psi$ from a Killing spinor $\phi$ as
\begin{equation}
\psi=\phi+\phi^{\varsigma} \nonumber
\end{equation}
where $\phi^{\varsigma}$ is also a Killing spinor obtained from the KS $\phi$ by the discrete transformation $\varsigma$:  we call it the \textit{Killing reversal map}. The important point in its definition is, if $\lambda$ is the Killing number of $\phi$ then $-\lambda$ is the Killing number of its Killing reversal. The relation of this reversal operation to charge conjugation, parity and time-reversal is studied in \cite{Acik II}.\\

Finally the brown arrow indicates the construction of Dirac spinor fields from the Parallel ones. It is known that in the ordinary Minkowski spacetime plane-wave Dirac solutions can be set up from parallel spinors with some additional constraints imposed for consistency \cite{Benn Tucker}-. A good account of the first quantised theory of the relativistic electron can be reached by this way and the complementary additional geometrical tools such as spinor products and spinorial Lie derivatives help respectively for achieving a complete spinor basis labeled by the eigenvalues of the quantum operators corresponding to energy and spin angular momentum representing their physical measurability. The intrinsic-spin degrees of freedom are shown to exist without the need for an externally applied magnetic field which to our view shows the physical nature of Lie transports of spinor fields in spacetime. However in the presence of high gravitational fields, Dirac solutions with non-compact (spacelike) supports such as plane waves are not physically well defined. In curved spacetimes the condition of global hyperbolicity ensures well behaved causal structures, so the flat spacetime quantum field theoretical tools could then be applied at least to scalar fields living in these globally hyperbolic spacetimes. This requires the elimination of refering to any plane wave basis which is possible due to some algebraic techniques. Hence a good warm up exercise can be reworking (in a second quantised manner) the relation between parallel spinors and Dirac spinors in Minkowski spacetime without any reference to plane waves. Since our main objects are Killing spinor fields in Riemannian spacetimes it is necessary to develop quantisation techniques for these fields. In canonical approaches the classical dynamics governed by the field equations should admit a well posed initial value formulation sensitive to the selection of a metric class for the background spacetime. We have found in \cite{Acik III} that Killing spinor fields can be used for constructing many other physical fields including the Klein-Gordon field, hence we aim to follow a reverse path for quantising KS fields in globally hyperbolic spacetimes by the help of their relation to scalar fields which admit a prescription for being quantised in this special class of backgrounds. If this can technically be done, the quantum dynamics of other fields obtainable from Killing spinors will be uncovered. This also will shed some light on the limited criteria for interpreting quantum field states as particles in curved spacetimes \cite{Wald,Fulling} because of our previous results relating smooth p-branes and Killing spinors.\\

 Although the theory of Killing spinors contain many tools suitable for beneficial use in curved spacetime quantum field theory, its contact with Killing vector fields opens another window in this area. An example at first glance can be given by generalising the quantum mechanics of the scalar field in Minkowski spacetime \cite{Tucker expectation} to a globally hyperbolic spacetime admitting a Killing spinor. In the latter case the magic will come from the fact that both the dynamical (scalar) field and the kinematical (time-like Kiling vector) field will be generated from the existent Killing spinor! This to our guess will interwine smooth spinorial differential geometry with the algebraic fabric of second quantisation in a different manner then usual.

\subsection{Frames}

Previously it was noted that the inner products of general spinor basis were $\mathbb{D}$-valued functions on $M$ (page $8$), but for a special kind of basis $\{b_i\}$ called the standard basis, it is possible to find a constant $\mathbb{D}$-matrix $\mathbf{C}$ either symmetric or skew such that
\begin{eqnarray}
(b_i,b_j)=C^{-1}_{ij} \nonumber
\end{eqnarray}
where $\mathbf{C} {\mathbf{\gamma}^{a}}^T \mathbf{C}^{-1}=-\mathbf{\gamma}^{a}$ and the transposition $T$ is similar to the adjoint involution $\mathcal{J}$ which is $\xi\eta$ in this case. Here the constant $\mathbb{D}$-valued gamma matrices $\{\boldsymbol{\gamma}^{a}\}$ satisfy $\boldsymbol{\gamma}^{a}\boldsymbol{\gamma}^{b}+\boldsymbol{\gamma}^{b}\boldsymbol{\gamma}^{a}=2g^{ab}\mathbf{1}$, the left Clifford action of orthonormal coframes on elements of standard spinor basis is given by $e^a b_i=\sum_{j} b_j \gamma^a_{ji}$. It is the last relation that identifies in some sense the standard spinor frames and $g$-orthonormal frames so their images under the spinor metric and spacetime metric respectively are all constants instead of functions! To gain more information from this relation, we emphasize the fact that putting the metric $g$ into orthonormal form $\eta_{ab}\,e^a\otimes e^b$ is understood as a reflection of the local validity of special relativity in Einstein's theory of classical gravitation. This choice is a visual illusion which makes one to think that gravitation has locally disappeared and special relativistic free float opportunity for test objects is created; of course we know that this is not the case since $\{{\Gamma_{ab}}^c\}\neq \{0\}$ where $\nabla_{X_a}X_b={\omega^c}_b(X_a)X_c={\Gamma_{ab}}^c X_c$. But we know from the \textit{local flatness theorem} in (pseudo-)Riemannian differential geometry in general and in Lorentzian differential geometry in special, that there exist coordinate patches trivialising the metric tensor and annihilating the connection coefficients at a point, a (geodesic) line or a higher dimensional submanifold. These are known to be the \textit{normal coordinates} for point supports and \textit{Fermi coordinates} for extended supports; as a consequence the selection of latter coordinates turns the previous visual appearance into reality at least to second order with respect to the metric. The improved tools such as Fermi vector fields, Darboux frames, Jacobi fields and geodesic deviation equation, exponential mapping and minimal varieties, variation vector fields and Lie derivatives \cite{Gray, Poisson, Morgan a, Hartley Tucker} are all of importance when analysing the geometry and dynamics of these substructures that are identified from a physical point of view with extended particles \cite{Dirac a, Dirac b, Lees, Gnadig et al, Barut, Barut Pavsic} . The existence of fermi (normal) coordinates can then be associated to the existence of local free-fall frames for extendons exhibiting 'geodesic' (minimal) motion.\\

 It is at this point that we want to ask the question whether one can gain further information about the dynamics of fermions in curved spacetimes by suitably applying the above methods to specific spinor frames coupled to the associated coframes. For example we know that the standard spinor frame $\{b_i\}$ is associated to the $g$-orhonormal coframe $\{e^a\}$ and the coupling is given by the previously mentioned left Clifford action $e^a b_i=\sum_{j} b_j \gamma^a_{ji}$. The next thing to do should be to deduce the spinor frame and the definitive coupling associated in the above sense to a Fermi coframe field. After gaining enough knowledge in this direction should one apply these tools to the extended models of the electron. Other applications may include free fall of electrons inside a superconductor immersed in a gravitational field \cite{Schiff Barnhill}, \cite{DeWitt}.

\section{Conclusion}

In the postface of Claude Chevalley's famous book \cite{Chevalley}, Jean Pierre Bourguignon quotes Roger Penrose by the words: ''\textit{We have not yet reached the age of \textbf{Spinormania}, but Roger Penrose \cite{Penrose Rindler} has seriously advocated that everything should be thought about in terms of spinors}''. Killing spinor programme can be seen as a part of \textit{Spinormania} and it mainly proposes that Killing spinors are the most primitive physical objects such that every other physical field or particle in one way or an other could be built up from them. However rather then trying to extract the constructive character of KSs during this work we focused on the web of relations between specific spinor fields and their homogeneous bilinears. But other newly deiscovered physical aspects of KSs and KS based objects are planned to be published elsewhere as part of our programme \cite{Acik II, Acik III}. Moreover a satisfactory knowledge of Dirac spinor GDCs are given here, but there seem to be many properties waiting to be digged out. Some hints about the so-called detailed approach are given throughout the text but its holistic structure will be worked out in an acompanying paper; since this will necessitate the tools related to complexification and realification then it will be possible to deeply connect our programme with the Twistor programme of Penrose. Penrose's main motivation for twistors was to consider all complex structures of spacetime at once for his Quantum Mechanical concerns (see the postface of \cite{Chevalley}).

\appendix
\section{Co-frame Independent Expansions of spinor bilinears}
We simply want to show the independence of equation (4) from the selection of a local co-frame field $\{e^{i}\}$. Since this set generates the whole local structure via Clifford multiplication and $\mathcal{F}(\mathbb{R})$-linearity, restricting ourselves only to degree $1$ will not deflect us from completeness. For the sake of generality, we work with coupled-bilinears and choose any two local co-frame fields $\{e^{i}\}$ and $\{E^{j}\}$. If we take the advantage of using the representatives of two well-known classes of co-frame fields, things will be much easier and useful. These are the natural co-frame field $\{dx^{\mu}\}$ and $g$-orthonormal co-frame field $\{e^{a}\}$ which are tied locally by the $GL(n,\mathcal{F}(\mathbb{R}))$ relations
\begin{eqnarray}
e^{a}={{\mathbf{e}}^{a}}_{\mu}(x)\,dx^{\mu}\:,\qquad g=g_{\mu\nu}(x)dx^{\mu}\otimes dx^{\nu}=\eta_{ab} \,e^{a}\otimes e^{b}
\end{eqnarray}
in the coordinate chart $x=(x^{\mu})$, where also the local forms of the metric tensor field with respect to the two co-frames are given. So as said above, the $1$-form part of the coupled-bilinear resolved in the $g$-orthonormal co-frame is
\begin{equation}
(\psi\bar{\phi})_{1}=(\phi,\,e_{a}^{\xi}\psi)e^{a}=(\phi,\,e_{a}\psi)e^{a}.\\ \nonumber
\end{equation}
Using the $\mathbb{R}$-matrix valued functions and $\mathcal{F}(\mathbb{R})$-bilinearity property of the spinor products we can write
\begin{eqnarray}
(\psi\bar{\phi})_{1}&=&(\phi,\,{{\mathbf{e}}_{a}}_{\mu}(x)\,dx^{\mu}\psi){{\mathbf{e}}^{a}}_{\nu}(x)\,dx^{\nu}=(\phi,dx^{\mu}\psi)\eta_{ab}{{\mathbf{e}}^{b}}_{\mu}(x){{\mathbf{e}}^{a}}_{\nu}(x)\,dx^{\nu}\\ \nonumber
&=&(\phi,dx^{\mu}\psi)g_{\mu\nu}(x)\,dx^{\nu}=(\phi,dx^{\mu}\psi)dx_{\mu}=(\phi,dx_{\mu}\psi)dx^{\mu},
\end{eqnarray}
which shows the independence from the chosen co-frame and thus completes the proof!

\section{The derivation of the canonical form for the elements of the kernel of Clifford contraction}
Let us write $e^{a}\otimes \pi_{X_{a}}=Id_{\Gamma(T^{*}M\otimes S)}=Id_{\Upsilon_{1} \oplus \Upsilon_{2}}$ so the twistor component map works as
\begin{eqnarray}
\Gamma(T^{*}M\otimes S)\stackrel{\widehat{pr}_{2}} \longrightarrow \Gamma \Upsilon_{2} &\stackrel{\pi_{X_{a}}}\longrightarrow& \Gamma S \\          \nonumber
\widetilde{X}\otimes \psi \mapsto \widehat{pr}_{2}(\widetilde{X}\otimes \psi)=e^{b}\otimes \widehat{\psi}_{b}^{X}&\mapsto& \pi_{X_{a}}(e^{b}\otimes \widehat{\psi}_{b}^{X}) =\widehat{\psi}_{a}^{X}\:.
\end{eqnarray}
We know that $\widehat{pr}_{2}(\widetilde{X}\otimes \psi)$ is $\widetilde{X}\otimes \psi-\frac{1}{n} e^{b}\otimes (e_{b}.\widetilde{X}.\psi)$ so lets calculate the image of this under the component map $\pi_{X_{a}}$ and reach to the spinor functional $\widehat{\psi}_{a}^{X}$ with its form as given before.
\begin{eqnarray}
\pi_{X_{a}}(\widetilde{X}\otimes \psi-\frac{1}{n} e^{b}\otimes e_{b}.\widetilde{X}.\psi)&=&\widetilde{X}(X_{a})\psi-\frac{1}{n}e^{b}(X_{a}) e_{b}.\widetilde{X}.\psi \\ \nonumber
&=&g(\widetilde{X},e_{a})\psi-\frac{1}{n} e_{a}.\widetilde{X}.\psi \\ \nonumber
&=&\frac{1}{n}(ng(\widetilde{X},e_{a}) - e_{a}.\widetilde{X}).\psi \\ \nonumber
&=&\frac{1}{n}(\frac{1}{2}(\widetilde{X}.e_{a}+e_{a}.\widetilde{X})n- e_{a}.\widetilde{X}).\psi \Rightarrow \\ \nonumber
\widehat{\psi}_{a}^{X}&=&\frac{1}{2n}((n-2)e_{a}.\widetilde{X}+n \widetilde{X}.e_{a}).\psi\:. \nonumber
\end{eqnarray}
The important property $e^{a}.\widehat{\psi}_{a}^{X}=0$ for all $a$ and vector fields $X$ can be seen from the ordinary Clifford identities $e^{a}. e_{a}=n$ and $e^{a}.\alpha_{p}.e_{a}=(n-2p)\alpha_{p}^{\eta}$ where $\alpha_{p} \in \Gamma \Lambda^{p}(M)$. There are two things of physical importance to say here:
\begin{enumerate}
  \item The property $e^{a}.\widehat{\psi}_{a}^{X}=0$ for a spinor valued $1$-form $e^{a}\otimes \widehat{\psi}_{a}^{X}$ corresponds to the vital tracelessness condition for the spin-$3/2$ tensor spinors $\Psi^{X}:=\widehat{\psi}_{a}^{X}\otimes e^{a}$ of SUGRA theories. Spin-$3/2$ tensor spinors are of the form $\Psi=b_{i}\otimes \psi^{i}=\psi_{a}\otimes e^{a}$; where $\{b_{i}\}$ is a standard spinor basis, $\{ e^{a}\}$ a $g$-orthonormal co-frame basis, the spinor $\psi_{a}=b_{i} \psi^{i}_{a}$ and the $1$-form $\psi^{i}=\psi^{i}_{a}e^{a}$. In fact $\Psi$ is independent of any chosen basis and for example if the natural co-frame $\{dx^{\mu}\}$ is selected then $\Psi=\psi_{\mu}\otimes dx^{\mu}$ and $\psi_{\mu}=\psi_{a}{e^a}_{\mu}=b_{i}\psi^{i}_{a}{e^a}_{\mu}$. Note that $\sharp\{e^{a}\}=n$, $\sharp\{ b_{i}\}\stackrel{\Bbbk=\mathbb{R}}{=}2^{\lfloor n/2\rfloor + \sigma}$; $\sigma=0,1$ and $\sharp\{ b_{i}\}\stackrel{\Bbbk=\mathbb{C}}{=}2^{\lfloor n/2\rfloor}$. The $\psi$ co-frame $\{\psi^i\}$ for $n>4$ has many redundant elements; this mathematical redundancy may have physical meanings especially in defining the motions of fermions or bosons in high gravitational fields. Spin-$3/2$ case requires a good understanding of representations of Clifford algebras on spinor valued $1$-forms; for higher spin fields a general and esthetic calculus for Clifford-valued $p$-forms together with their left action on spinor-valued homogeneous degree-forms could be find e.g. in \cite{Benn Tucker} (see also \cite{Benn Al Saad Tucker}).

  \item The first Clifford identity mentioned above i.e. $e^{a}. e_{a}=n$ appears almost every where in calculations and could be used as a key identity for unifying four fundamental forces of physics. This is used to define a new gauge connection from the Lagrangian density for a massive Dirac spinor field minimally coupled to the electromagnetic field in curved spacetime. This theory invented by Chisholm and Farwell in 1987 is called the \textit{spin gauge theory} and it has the power to unify the four forces without the complex machinery of spontaneous symmetry breaking and Higgs mechanism \cite{Chisholm Farwell, Lyle}. In contrast to the standard model of particle physics, these theories could predict the particle masses and the major argument against them is that they are highly mathematical. This is like saying that the three dimensional vector calculus is not suitable for formulating Newtonian mechanics or Maxwellian electrodynamics, because it is more mathematical than desired! One should as well remember Einstein's objection to Minkowski's \textit{'Lorentz covariant (spacetime) formulation of Special Relativity'}, that is treating space and time strangely in an equal footing!
\end{enumerate}

\section{Dirac spinors and their generalised Dirac currents}
Let $\psi$ be a Dirac spinor, then it satisfies the Dirac equation
\begin{eqnarray}
\displaystyle{\not}D \psi=m \psi. \nonumber
\end{eqnarray}
We want to learn the equations satisfied by the homogeneous bilinears $(\psi \overline{\psi})_{p}$. From the compatibility of the spinor connection with invariant inner products we first write
\begin{eqnarray}
\nabla_{X_{a}}(\psi \overline{\psi})_{p}=(\nabla_{X_{a}}\psi \overline{\psi})_{p}+(\psi \overline{\nabla_{X_{a}}\psi})_{p}. \nonumber
\end{eqnarray}
and the other steps include calculating the exterior derivative and the co-derivative of this $p$-form. The former one follows as
\begin{eqnarray}
d(\psi \overline{\psi})_{p}&=&e^{a}\wedge(\nabla_{X_{a}}\psi \overline{\psi})_{p}+e^{a}\wedge(\psi \overline{\nabla_{X_{a}}\psi})_{p} \\ \nonumber
&=&(e^{a}\wedge(\nabla_{X_{a}}\psi \overline{\psi}))_{p+1}+(e^{a}\wedge(\psi \overline{\nabla_{X_{a}}\psi}))_{p+1}. \nonumber
\end{eqnarray}
Remembering the properties of Clifford multiplication
\begin{eqnarray}
A.\Phi+\Phi^{\eta}.A&=&2 A\wedge \Phi, \\
A.\Phi-\Phi^{\eta}.A&=&2i_{\widetilde{A}}\Phi,
\end{eqnarray}
here $A$ is a $1$-form and $\Phi$ is any (possibly inhomogeneous) Clifford form; then we can write $(C1)$ as
\begin{eqnarray}
d(\psi \overline{\psi})_{p}&=&\frac{1}{2}[e^{a}.(\nabla_{X_{a}}\psi \overline{\psi})+(\nabla_{X_{a}}\psi \overline{\psi})^{\eta}.e^{a}+e^{a}.(\psi \overline{\nabla_{X_{a}}\psi})+(\psi \overline{\nabla_{X_{a}}\psi})^{\eta}.e^{a}]_{p+1} \\ \nonumber
&=&\frac{1}{2}[m(\psi \overline{\psi})_{p+1}-(-1)^{p+1}(\nabla_{X_{a}}\psi \overline{\psi}.e^{a})_{p+1}+(e^{a}.\psi \overline{\nabla_{X_{a}}\psi})_{p+1}-(-1)^{p+1}(\psi \overline{\nabla_{X_{a}}\psi}.e^{a})_{p+1}] \\ \nonumber
&=&\frac{1}{2}[m(\psi \overline{\psi})_{p+1}-(-1)^{p+1}(\nabla_{X_{a}}\psi \overline{\psi}.e^{a})_{p+1}+(e^{a}.\psi \overline{\nabla_{X_{a}}\psi})_{p+1}-(-1)^{p+1}(\psi \overline{(e^{a})^{\mathcal{J}}\nabla_{X_{a}}\psi})_{p+1}] \\ \nonumber
&=&\frac{1}{2}[(m\pm(-1)^{p+1} m^{j_{\Bbbk}})(\psi \overline{\psi})_{p+1}-(-1)^{p+1}(\nabla_{X_{a}}\psi \overline{\psi}.e^{a})_{p+1}+(e^{a}.\psi \overline{\nabla_{X_{a}}\psi})_{p+1} ].  \nonumber
\end{eqnarray}
In the above equations we have used the associativity of Clifford product, the effect $(e^{a})^{\mathcal{J}}=\pm e^{a}\, (+ $for$ \mathcal{J}\simeq \xi, \xi^{*})$ and the fact that $m$ can be real or pure imaginary which causes us use $j_{\Bbbk}$ which is induced on $\Bbbk=\mathbb{R},\mathbb{C}$ from $j$ on $\mathbb{D}$ . Likewise for the co-differential a similar calculation goes as
\begin{eqnarray}
\delta(\psi \overline{\psi})_{p} &=&-[i_{X^{a}}(\nabla_{X_{a}}\psi \overline{\psi})+i_{X^{a}}(\psi \overline{\nabla_{X_{a}}\psi})]_{p-1} \\ \nonumber
&=&-\frac{1}{2}[e^{a}.(\nabla_{X_{a}}\psi \overline{\psi})-(\nabla_{X_{a}}\psi \overline{\psi})^{\eta}.e^{a}+e^{a}.(\psi \overline{\nabla_{X_{a}}\psi})-(\psi \overline{\nabla_{X_{a}}\psi})^{\eta}.e^{a}]_{p-1} \\ \nonumber
&=&-\frac{1}{2}[(m\pm(-1)^{p-1} m^{j_{\Bbbk}})(\psi \overline{\psi})_{p-1}+(-1)^{p-1}(\nabla_{X_{a}}\psi \overline{\psi}.e^{a})_{p-1}+(e^{a}.\psi \overline{\nabla_{X_{a}}\psi})_{p-1}].  \nonumber
\end{eqnarray}
Another step can be taken by the observation $(\nabla_{X_{a}}\psi \overline{\psi}.e^{a})_{q}^{\mathcal{J}}=\varepsilon \delta(\mathcal{J})(e^{a}.\psi \overline{\nabla_{X_{a}}\psi})_{q}$ where $(e^{a})^{\mathcal{J}}:=\delta(\mathcal{J})e^{a}$, $\varepsilon$ is the symmetry indicator of the spinor inner-product and $\delta(\mathcal{J})=\delta_{[\mathcal{J}],[\xi]}-\delta_{[\mathcal{J}],[\xi\eta]}+\delta_{[\mathcal{J}],[\xi^{*}]}-\delta_{[\mathcal{J}],[\xi\eta^{*}]}$
so
\begin{eqnarray}
d(\psi \overline{\psi})_{p}=\frac{1}{2}[(m+(-1)^{p+1}\delta(\mathcal{J}) m^{j_{\Bbbk}})(\psi \overline{\psi})_{p+1}+(1-(-1)^{p+1}\epsilon\delta(\mathcal{J})\mathcal{J})(e^{a}.\psi \overline{\nabla_{X_{a}}\psi})_{p+1}]  \nonumber
\end{eqnarray}
and
\begin{eqnarray}
\delta(\psi \overline{\psi})_{p}=-\frac{1}{2}[(m+(-1)^{p-1}\delta(\mathcal{J}) m^{j_{\Bbbk}})(\psi \overline{\psi})_{p-1}+(1+(-1)^{p-1}\epsilon\delta(\mathcal{J})\mathcal{J})(e^{a}.\psi \overline{\nabla_{X_{a}}\psi})_{p-1}].  \nonumber
\end{eqnarray}
Another simplifying identity that we call the \textit{key identity} $$(e^{a}.\psi \overline{\nabla_{X_{a}}\psi})_{q}=(e^{a}.\nabla_{X_{a}}(\psi \overline{\psi}))_{q}-(e^{a}.(\nabla_{X_{a}}\psi) \overline{\psi})_{q}=(\displaystyle{\not}d(\psi \overline{\psi}))_{q}-m(\psi \overline{\psi})_{q}$$ is given then everything can be written in terms of homogeneous singled-bilinears after the definitions $\mathbf{m}(p,j_{\Bbbk}):=m+(-1)^{p-1}\delta(\mathcal{J}) m^{j_{\Bbbk}}$ and $\Delta_{\pm}(p,\mathcal{J}):=(1\pm(-1)^{p+1}\epsilon\delta(\mathcal{J})\mathcal{J})$ as
\begin{eqnarray}
(2-\Delta_{-})d(\psi \overline{\psi})_{p}=(\mathbf{m}-m\Delta_{-})(\psi \overline{\psi})_{p+1}-\Delta_{-} \delta(\psi \overline{\psi})_{p+2}  \nonumber
\end{eqnarray}
and
\begin{eqnarray}
-(2-\Delta_{+})\delta(\psi \overline{\psi})_{p}=(\mathbf{m}+m\Delta_{+})(\psi \overline{\psi})_{p-1}+\Delta_{+}d(\psi \overline{\psi})_{p-2}.  \nonumber
\end{eqnarray}
If the key identity is zero we reach the \textit{key constraint}
$$(e^{a}.\psi \overline{\nabla_{X_{a}}\psi})_{q}=0$$
generating the massive \textit{K\"{a}hler equation}
$$(\displaystyle{\not}d(\psi \overline{\psi}))_{q}=m(\psi \overline{\psi})_{q}\:;\quad\exists q\;or\:\forall q$$
in this context.

\acknowledgments
I would like to thank my friend and colleague \"{U}mit Ertem for helpful discussions and for his careful proofreading; and I am indebted to Robin W. Tucker for providing me a copy of his paper \cite{Tucker expectation} and for his inspiring style for doing physics.

 \end{document}